\documentclass[aps,prd,twocolumn,showpacs,preprintnumbers,nofootinbib,amsmath,amssymb]{revtex4}

\usepackage{graphicx}% Include figure files
\usepackage{dcolumn}% Align table columns on decimal point
\usepackage{bm}% bold math
\usepackage{amsmath}
\usepackage[normalem]{ulem}

\newcommand{\tr}{\ensuremath{\mathrm{Tr}}}

\newcommand{\mbf}[1]{\ensuremath{\boldsymbol{#1}}}
\DeclareRobustCommand{\Eqref}[1]{\eqref{#1}}

\begin{document}

\title{
Magnetic susceptibility and equation of state of $N_f = 2+1$ QCD
with physical quark masses
}
\author{Claudio Bonati}
\email{bonati@df.unipi.it}
\affiliation{
Dipartimento di Fisica dell'Universit\`a
di Pisa and INFN - Sezione di Pisa,\\ Largo Pontecorvo 3, I-56127 Pisa, Italy}

\author{Massimo D'Elia}
\email{delia@df.unipi.it}
\affiliation{
Dipartimento di Fisica dell'Universit\`a
di Pisa and INFN - Sezione di Pisa,\\ Largo Pontecorvo 3, I-56127 Pisa, Italy}

\author{Marco Mariti}
\email{mariti@df.unipi.it}
\affiliation{
Dipartimento di Fisica dell'Universit\`a
di Pisa and INFN - Sezione di Pisa,\\ Largo Pontecorvo 3, I-56127 Pisa, Italy}

\author{Francesco Negro}
\email{fnegro@ge.infn.it}
\affiliation{Dipartimento di Fisica dell'Universit\`a
di Genova and INFN - Sezione di Genova,\\
 Via Dodecaneso 33, I-16146 Genova, Italy}
\altaffiliation[Now at: ]{INFN - Sezione di Pisa, Largo Pontecorvo 3, I-56127 Pisa, Italy}

\author{Francesco Sanfilippo}
\email{sanfilippo@soton.ac.uk}
\affiliation{Laboratoire de Physique Th\'eorique (Bat. 210) 
  Universit\'e ParisSUD, F-91405 Orsay-Cedex, France}
\altaffiliation[Now at: ]{School of physics and astronomy, University of Southampton, 
  SO17 1BJ Southampton, United Kindgdom}

\date{\today}% It is always \today, today,

\begin{abstract}
We determine the free energy of strongly interacting matter
as a function of an applied constant and uniform magnetic field.
We consider $N_f = 2+1$ QCD with physical quark masses,
discretized on a lattice by
stout improved staggered fermions and a tree level improved 
Symanzik pure gauge action, 
and explore three different lattice spacings. 
For magnetic fields of the order of those 
produced in non-central heavy ion collisions ($eB \sim 0.1$ GeV$^2$)
strongly interacting matter behaves like a medium with a linear
response, 
and is paramagnetic both above 
and below the deconfinement transition, with a susceptibility
which steeply rises in the deconfined phase. 
We compute the equation of state,
showing that the relative increase in the pressure due to the magnetic
field gets larger around the transition, and of the order of 
10\% for $eB \sim 0.1$ GeV$^2$.
\end{abstract}

\pacs{12.38.Aw, 11.15.Ha,12.38.Gc,12.38.Mh}
\maketitle

\section{Introduction}
\label{intro}

Strong interactions present some of the most interesting 
open issues within the Standard Model of particle physics.
Quarks and gluons, the elementary colored degrees of freedom of 
Quantum Chromodynamics (QCD), appear to be confined in ordinary
matter, due to a mechanism of color confinement 
which is not yet fully understood. On the other hand, 
theoretical arguments~\cite{cabibbo} and lattice QCD simulations
predict the emergence of new phases, in which quarks and gluons
are deconfined, when strongly interacting matter is put under 
extreme conditions, characterized by a high temperature $T$ and/or 
a high baryon density.
A strongly interacting plasma of quarks and gluons (QGP), in 
particular, is the new
state of matter which is thought to have filled our Universe in its
early stages, and which is hunted for by experiments exploiting 
ultra-relativistic heavy ion collisions. 

We are interested here in the properties
of this new ``material'', and of strongly interacting matter
more in general, when it is placed in the 
presence of strong magnetic fields. The issue is fundamental,
both theoretically and phenomenologically, since in some heavy ion 
collisions one has the highest magnetic fields ever created 
in a laboratory~\cite{hi1,hi2,hi3,hi4}, reaching 
up to $O(10^{15})$ Tesla (i.e. $e B$ of  $O(0.1)$ GeV$^2$), 
and even larger fields may have 
been created in the early stages of the Universe~\cite{vacha,grarub}.
That justifies the recent theoretical investigations on the 
subject (for comprehensive reviews on various topics, 
see Ref.~\cite{lecnotmag}).
In the non-perturbative regime of the theory, this problem 
can be conveniently approached by lattice QCD 
simulations~\cite{buivid, buivid1, buivid2, buivid3, blum, reg1, 
reg2, levkova, cffmm, DEN, reg0, demusa, 
thetaeff, ilgen, ilgen2, reg3, reg4, bari}.
The question regards, in particular, the dependence of the free energy
density on a background field $B$, which contains information
on the equation of state
as a function of $B$
and reveals whether the system is diamagnetic 
or paramagnetic.

The issue is well posed and has been approached
by a number of analytic~\cite{Ioffe:1983ju, Belyaev:1984ic, EPS, Braun:2002en, 
Kim:2004hd, Rohrwild:2007yt, Goeke:2007nc, Bergman:2008sg, agasian08, Gorsky:2009ma, 
Frasca:2011zn, F1, F2, F3, endrodihrg, Nam:2013wja, Anber:2013tra, Steinert:2013fza} 
and recent lattice studies~\cite{buivid,reg1,reg2,cffmm,levkova,reg3,reg4}.
In a typical lattice setup,
the need for magnetic flux quantization
introduces a few technical difficulties when one tries to compute
the derivatives of the free energy density with respect to $B$.
In Ref.~\cite{cffmm}, we have proposed 
to overcome such difficulties by determining
free energy finite differences, in place of its derivatives, and we 
have applied this method to the determination of the magnetic susceptibility
for $N_f = 2$ QCD, in the standard rooted staggered discretization,
with pion masses going down to 200~MeV. 
Results have provided evidence for a weak magnetic activity in the 
confined phase and for the emergence of strong paramagnetism in 
the deconfined, Quark-Gluon Plasma phase; such evidence has been
confirmed by Refs.~\cite{levkova,reg3,reg4}.

The purpose of the present study is to improve on our original 
determination in several aspects. First of all, we investigate
$N_f = 2+1$ QCD with physical quark masses and three different 
lattice spacings, in order to check for discretization effects.
In this way, we get control over the two major sources of systematic error
and provide information which is of direct phenomenological relevance.
Second, we will extend the range of temperatures and try to better
resolve the region around deconfinement, in order to understand if
paramagnetism, or diamagnetism instead, takes place below $T_c$, and
to obtain more information about the high-$T$ behavior of the susceptibility.

The paper is organized as follows. In Section~\ref{method} we review
the method adopted to determine the renormalized free energy density 
as a function of the background field. In Section~\ref{setup}
we illustrate the lattice discretization and the numerical
setup adopted. In Section~\ref{results} we present and discuss our results.
Finally, in Section~\ref{summary}, we draw our conclusions and 
discuss future perspectives.

\section{The Method}
\label{method}

The main focus of our study is on the dependence of the free energy density
$f$ of QCD on an external constant and uniform 
magnetic field $\mbf{B}$. 
Usually, the dependence is given in terms
of the derivatives of $f$ with respect to $B$,
like the magnetization (first derivative) or the
susceptibility (second derivative), which are
more easily measureable quantities.
However, in a lattice setup, 
taking such derivatives is not trivial, due to field
quantization.

Indeed, in order to 
minimize finite size effects,
one usually works
on a compact manifold, like a 3D torus.
Consistency conditions
then require that the magnetic field flux through each closed 
surface either vanish or be quantized in units of
$2 \pi /q$, where $q$ is the elementary electric charge
of the particles which populate the system.
In the case of quarks ($q = |e|/3$)
moving in a uniform field 
$\mbf{B} = B\ \mbf{\hat z}$ on a 3D torus
one has~\cite{bound1,bound2,bound3,wiese}
\begin{equation}\label{bquant}
|e| B = {6 \pi b}/{(\ell_x \ell_y)} 
\end{equation}
where $\ell_x$, $\ell_y$ are the torus extensions 
in the $x, y$ directions and $b$ must take integer values.

A few approaches have been devised to get around this problem.
In Refs.~\cite{reg2,reg4}, the $B$-dependent part of the free energy density
has been related to the pressure anisotropy, whose determination requires
the knowledge of some perturbative coefficients. In Ref.~\cite{levkova}
the lattice torus has been divided in two identical regions
where the magnetic field takes uniform but opposite values, thus ensuring
an overall zero magnetic flux without any need for quantization. That allows
to take $B$-derivatives as usual, even if at the price of introducing
interface effects at the boundary between the two regions,
which are expected to be negligible in the large volume limit.
In Ref.~\cite{cffmm}, instead, we have developed a method which allows
to compute the $B$-dependent part of $f$,
\begin{equation}
\Delta f(B,T) = 
- \frac{T}{V} \log \left( \frac{Z(B,T,V)}{Z(0,T,V)} \right)
\label{freediff}
\end{equation}
where $Z = \exp(-F/T)$ 
is the partition function of the system
and $V$ is the spatial volume. As it usually happens
when trying to evaluate the ratio of two partition 
functions~\cite{umbrella},
the method is computationally demanding, however it is well
defined and feasible.

The idea~\cite{cffmm} is to obtain the finite free energy differences
$f(B_2) - f(B_1) = f(b_2) - f(b_1)$, where $b_1$ and $b_2$ are integers,
by integrating the derivative of a suitable extension
of the function $f(b)$ to real values of $b$, 
\begin{equation}\label{intfb}
f(b_2) - f(b_1) = \int_{b_1}^{b_2} \frac{\partial f(b)}{\partial b} \mathrm{d} b\, .
\end{equation}
The operation is well defined as long as the interpolating
function $f(b)$ is continuous and differentiable, and
reduces the problem of determining the ratio of two partition functions
to a determination of a standard observable, $\partial f / \partial b$.
It must be stressed that this observable has no direct relation
with the magnetization of the original theory, even for 
integer values of $b$, since it is just the derivative of the
interpolating function. 

In practice, $f(b)$ can correspond 
to any distribution of magnetic field which interpolates between 
quantized values. In our realization,
the magnetic field will be uniform over the whole lattice apart
from a single plaquette, pierced by a sort of Dirac string which 
brings the excess flux away (see the next section for its explicit form). 
Of course the final result, $f(b_2) -f(b_1)$, is interpolation independent;
that has been also numerically verified in Ref.~\cite{cffmm}.

Once $\Delta f$ has been determined, one must take care of getting
rid of ultraviolet (UV) divergences,
since they contain $B$-dependent contributions
which do not cancel when taking the difference
$\Delta f$, and must
be properly subtracted.
Since we are interested in the magnetic properties of the thermal medium,
the most natural prescription is to subtract vacuum (i.e. $T = 0$)
contributions~\cite{cffmm}:
\begin{equation}\label{subtr}
\Delta f_R (B,T) = 
\Delta f(B,T) - \Delta f(B,0) \, ,
\end{equation}
where it is assumed that both terms on the right hand side are computed
at the same value of the UV cutoff (lattice spacing).
No further divergences are present in Eq.~(\ref{subtr}), since 
$B$-dependent divergences cannot depend also on $T$, apart 
from possible finite terms which vanish in the continuum limit
(see, e.g., the discussion in Refs.~\cite{reg0,reg2}).

The small field behavior of $\Delta f_R$ 
will give access to 
the magnetic susceptibility of the medium. 
Indeed, once vacuum contributions have been subtracted,
one has the relation 
\begin{equation}
\Delta f_R = -\int \mbf{\mathcal{M}}\cdot\mathrm{d}\mbf{B}\, ,
\end{equation}
where $\mbf{\mathcal{M}}$ is the magnetization of 
medium. Assuming that the medium is linear, homogeneous and 
isotropic, one has 
$\mbf{\mathcal{M}} = \tilde\chi \mbf{B}/\mu_0$, where
$\tilde\chi$ is the magnetic susceptibility in SI units, so that
\begin{equation}
\Delta f_R =-\frac{\tilde\chi}{\mu_0}\int \mbf{B}\cdot\mathrm{d}\mbf{B}=-\frac{\tilde\chi}{2\mu_0} B^2\
\label{intfree2} \, .
\end{equation}

The field $\mbf{B}$ appearing in Eq.~(\ref{intfree2})
is the total field acting on the medium. In our numerical setup,
the dynamics of electromagnetic fields will be 
quenched, so that there is no backreaction from the medium
itself, i.e. the magnetization $\mbf{\mathcal{M}}$ does not change
the value of the magnetic field. Therefore, $\mbf{B}$
coincides with the applied external field. While that does not introduce any 
systematic uncertainty in the determination of $\tilde\chi$,
which is the physical quantity that we will determine,
one should consider that the actual change in the free energy density 
of a real medium,
i.e. capable of producing magnetic fields by itself, will
be different. Indeed, if we 
introduce the auxiliary field 
$\mbf{H} = \mbf{B}/\mu_0 - \mbf{\mathcal{M}}$,
which is generated by external currents only, then
$$
\Delta f_R = -\frac{\mu_0 \chi (1 + \chi)}{2} H^2\
$$
where $\chi$ is the other standard definition of 
magnetic susceptibility,
$\mbf{\mathcal{M}}=\chi\mbf{H}$
($\chi = \tilde \chi / (1 - \tilde \chi)$).
A simple comparison of the two expressions shows that
the backreaction of the medium leads to a change
of $\Delta f_R$ by a factor $1/(1 - \tilde\chi)^2$.
However, considering the typical values of 
$\tilde\chi$ that we will show in Section~\ref{results}
($\tilde\chi \sim O(10^{-3})$), this correction turns out to be negligible.

In the following it will be convenient to express
$\Delta f_R$ also in the alternative form
\begin{equation}
\Delta f_R = -\frac{\hat\chi}{2} (e B)^2\
\label{intfree3} \, ,
\end{equation}
which is particularly useful when working in natural units.

\section{Numerical Setup}
\label{setup}

In this section we discuss our numerical setup for the discretization of
$N_f=2+1$ quark flavors in the presence of an external magnetic field,
with isospin symmetry broken only by the different electric charges.
The external electromagnetic field enters the QCD Lagrangian through quark 
covariant derivatives $D_\mu = \partial_\mu+igA_\mu^aT^a+iq_fA_\mu$,
where $A_\mu$ is the abelian gauge four potential and $q_f$ is the electric charge of the quark.
On the lattice, $SU(3)$ covariant derivatives are written in term of the parallel transport
$U_{i;\,\mu}$ ($i$ is the position and $\mu$ the direction) and the introduction of the 
abelian gauge field amounts to add also an $U(1)$ phase: $U_{i;\,\mu}\to u_{i;\,\mu}U_{i;\,\mu}$.

The euclidean partition function of the discretized
theory in the presence of a magnetic field is expressed as
\begin{eqnarray}
\label{partfunc}
\mathcal{Z}(B) &=& \int \!\mathcal{D}U \,e^{-\mathcal{S}_{Y\!M}} \!\!\!\!\prod_{f=u,\,d,\,s} \!\!\! \det{({D^{f}_{\textnormal{st}}[B]})^{1/4}}, \\
\label{tlsyact}
\mathcal{S}_{Y\!M}&=& - \frac{\beta}{3}\sum_{i, \mu \neq \nu} \left( \frac{c_0}{2}
W^{1\!\times \! 1}_{i;\,\mu\nu} +c_1W^{1\!\times \! 2}_{i;\,\mu\nu} \right), \\
\label{fermmatrix}
(D^f_{\textnormal{st}})_{i,\,j}&=&am_f \delta_{i,\,j}+\!\!\sum_{\nu=1}^{4}\frac{\eta_{i;\,\nu}}{2} 
\left(u^f_{i;\,\nu}U^{(2)}_{i;\,\nu}\delta_{i,j-\hat{\nu}} \right. \nonumber\\
&-&\left. u^{f*}_{i-\hat\nu;\,\nu}U^{(2)\dagger}_{i-\hat\nu;\,\nu}\delta_{i,j+\hat\nu}  \right)
\end{eqnarray}
where $\mathcal{D}U$ is the functional integration over the non-abelian $SU(3)$ gauge link variables,
$\mathcal{S}_{Y\!M}$ is the pure gauge action and 
$D^f_{\textnormal{st}}$ is the stout-improved staggered Dirac operator.
No integration is performed on the $U(1)$ link
variables, i.e. the electromagnetic degrees of freedom are quenched.

The action used for the gauge fields ($\mathcal{S}_{Y\!M}$) is the tree level improved Symanzik action
~\cite{weisz,curci},
which involves not only the real part of the trace of the 
standard $1\!\times \! 1$ square loops ($W^{1\!\times \! 1}_{i;\,\mu\nu}$)
but also that of 
the $1\!\times \!2$ rectangles ($W^{1\!\times \! 2}_{i;\,\mu\nu}$). 
The coefficients present in \eqref{tlsyact} 
are set to the values $c_0=5/3$ and $c_1=-1/12$.

In the continuum, the electromagnetic four potential giving rise to 
a uniform magnetic field $\mbf{B}=B\hat z$ can be written,
apart from constant terms, 
in the form 
$A_y=B x$ and $A_\mu=0$ for $\mu=t,\,x,\,z$. The discretization 
of such a field on a torus is not completely trivial and a possible choice for the $U(1)$ phases on the lattice is
\begin{eqnarray}
\label{bfieldy}
&& u^f_{i;\,y}=e^{i a^2 q_f B (i_x-L_x\Theta(i_x-L_x/2))}, \\
\label{bfieldx}
&&{u^f_{i;\,x}|}_{i_x=L_x/2}=e^{-ia^2 q_f L_x B (i_y-L_y\Theta(i_y-L_y/2)}
\end{eqnarray}
with $u^f_{i;\,\mu}=1$ elsewhere.
Here $\Theta(x)$ is the Heaviside step function and the term in Eq.~(\ref{bfieldx}) is required to 
guarantee the smoothness of the background field and the gauge invariance of the fermion action
(see, e.g., the discussion in Ref.~\cite{wiese}).

The particular form of the $U(1)$ field given above will be kept also 
for non-integer values of $b$: that fixes our choice for the 
free energy density interpolating 
between quantized values of the magnetic field. 
This choice is different from other standard discretizations 
found in the literature 
(see, e.g., Ref.~\cite{wiese}), and in particular from the one used 
in previous studies~\cite{DEN, cffmm, demusa, thetaeff}. The  
difference consists in a simple shift of the Dirac string from
one corner to the middle of the lattice. 
With the present choice, 
the variation and the magnitude of the phases is minimized,
leading to a significant improvement of   
the signal to noise ratio, as noted in Ref.~\cite{levkova}.

In the fermionic sector we have adopted the stout link 
smearing improvement~\cite{morning}
to reduce the effects of the finite lattice spacing and, in particular,
to reduce the taste symmetry violations (see Ref.~\cite{bazavov} 
for a comparison of the effectiveness of the various approaches).
The usual rooting trick is used in Eq.~(\ref{partfunc}) to eliminate the residual $4$ degeneracy which is
present in the staggered fermion spectrum.

The improved Dirac matrix 
$D^f_{\textnormal{st}}$
is built up by means of the two times stout smeared links $U^{(2)}_{i;\,\mu}$, which are
recursively defined by
(see \cite{morning})
\begin{equation}
\begin{aligned}
\label{stout}
& C^{(n)}_{i;\,\mu} = \sum_{\mu\neq\mu} \rho_{\mu\nu}\left( U^{(n)}_{i;\,\nu}U^{(n)}_{i+\hat\nu;\,\mu}
U_{i+\hat\nu;\,\nu}^{(n)\dagger}  \right.  \\
&\hspace{1cm} + \left. U_{i-\hat\nu;\,\nu}^{(n)\dagger}U_{i-\hat\nu;\,\mu}^{(n)}U_{i-\hat\nu+\hat\mu;\,\nu}^{(n)}\right), \\ 
&\Xi_{i;\,\mu}^{(n)} = U_{i;\,\mu}^{(n)}C_{i;\,\mu}^{(n)\dagger}-C_{i;\,\mu}^{(n)}U_{i;\,\mu}^{(n)\dagger},  \\
& Q_{i;\,\mu}^{(n)} = \frac{i}{2}\Xi_{i;\,\mu}^{(n)}-\frac{i}{2N_c}\tr\left(\Xi_{i;\,\mu}^{(n)}\right),  \\
& U_{i;\,\mu}^{(n+1)} = \exp \left( i Q_{i;\,\mu}^{(n)}\right)U_{i;\,\mu}^{(n)}\ ,
\end{aligned}
\end{equation}
where the $U_\mu^{(0)}$'s are the original integration link variables, 
which are used to compute the pure gauge action in Eq.~(\ref{tlsyact}).
In our simulations we adopted the isotropic smearing parameters $\rho_{\mu\nu}=0.15\,\delta_{\mu\nu}$.
The analyticity of the stout smearing procedure enables to straightforwardly implement the hybrid Monte Carlo 
update for the improved fermionic action, following Ref.~\cite{morning}.

Let us explicitly notice that, similarly to  
Refs.~\cite{reg0,reg1,reg2},
in our simulations the stout smearing is applied only to $SU(3)$ links, while
the external $U(1)$ phases are left untouched. 
See Ref.~\cite{lecnotmag2} for a discussion on this point.

\begin{figure}[t!]
\includegraphics[width=0.9\columnwidth, clip]{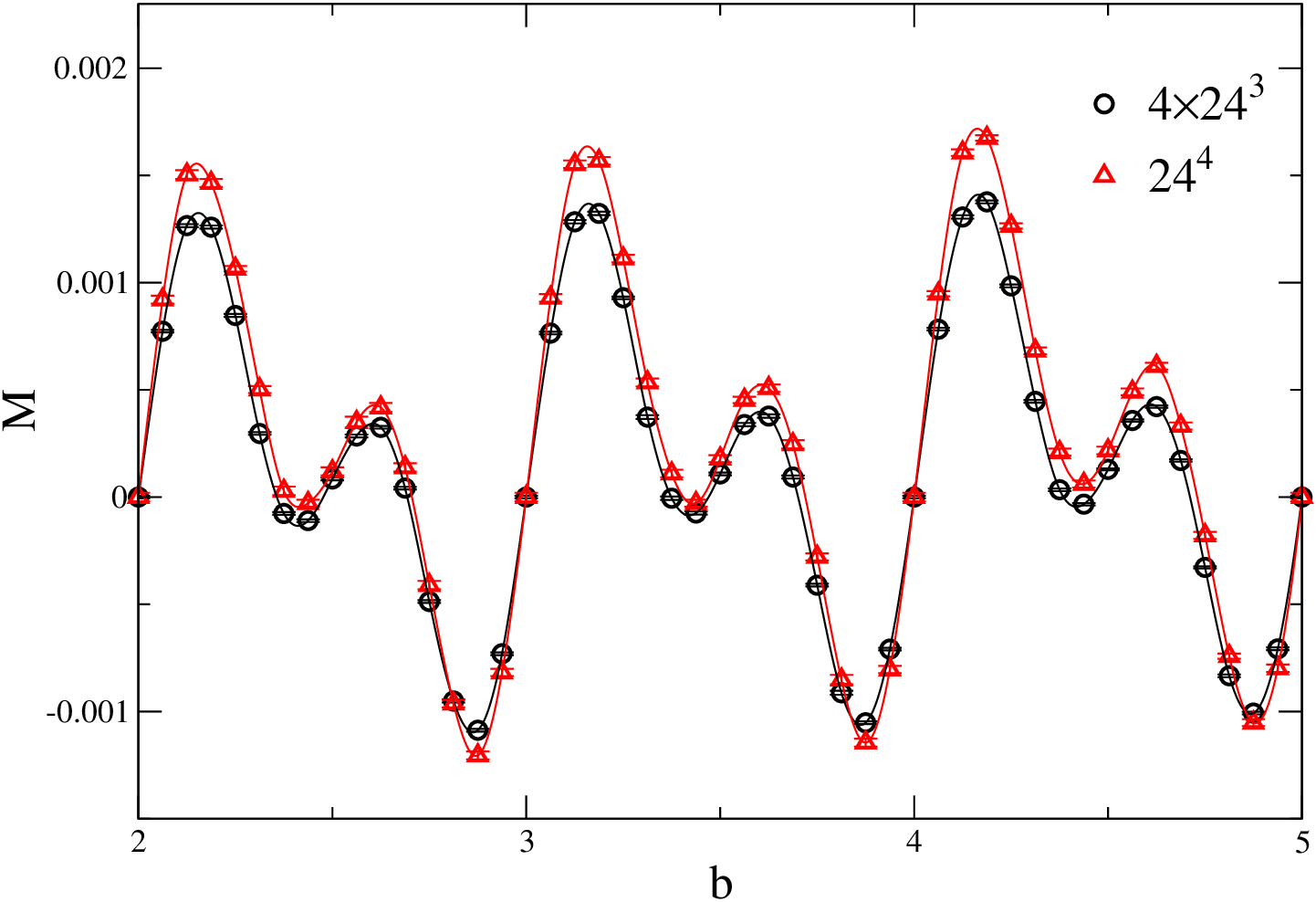}
\caption{$M$ computed on $24^4$ and $24^3 \times 4$ lattices, 
with $a \simeq 0.2173$\,fm.
The continuous lines correspond to third order spline interpolations.}
\label{M_fig}
\end{figure}

\begin{table}[t]
\centering
\begin{tabular}{ |c|c|c|c|c| }
\hline
$L_s$ & $a (\textnormal{fm})$ & $\beta$  & $am_{u/d}$ &$am_s$  \\
\hline
24 & 0.2173(4) & 3.55  &  0.003636 & 0.1020 \\
32 & 0.1535(3) & 3.67  &  0.002270 & 0.0639 \\
40 & 0.1249(3) & 3.75  &  0.001787 & 0.0503 \\
\hline
\end{tabular}
\caption{Simulation parameters. The bare coupling $\beta$ and the quark masses $am_{u/d}$ 
and $am_s$ are chosen according to what reported in Ref.~\cite{tcwup1,befjkkrs}, and correspond 
to a physical pion mass. The errors reported for the lattice spacing are the statistical 
ones, the systematical error is estimated to be about $2\%$ (see Ref.~\cite{tcwup1}).}
\label{param}
\end{table}

We have performed simulations at the physical value of the pion mass, 
$m_\pi \sim 135$ MeV, using the bare parameters $\beta$, $am_{u,d}$ and $am_s$ 
($m_s/m_{u,d}$ is fixed to its physical value, 28.15)
taken from Ref.~\cite{befjkkrs} and reported in 
Table~\ref{param},
which correspond to a line of constant physics
at three different values of the lattice spacing $a$.

The number of lattice sites in the spatial direction has been 
chosen so as to maintain a spatial extent around
5 fm for all values of  $a$.
At fixed bare parameters, the temperature of the system has 
been changed by varying the temporal extent of the lattice;
in our simulations we explored the temperature range 
$T\sim90-400~\textnormal{MeV}$, while our reference $T = 0$ symmetric 
lattices, used to subtract the vacuum contribution, correspond to 
temperatures below 40 MeV. As we discuss later, in 
Section~\ref{discussion}, the fact that the subtraction point
is at a low but finite $T$ does not introduce any appreciable 
systematics, due to the rapid convergence to zero of the 
susceptibility in the low $T$ region (exponentially in $1/T$).

Finally, the integrand appearing in Eq.~(\ref{intfb}),
$\partial f /\partial b$, has the following expression:
\begin{equation}
M \equiv a^4 \frac{\partial f}{\partial b} = \frac{1}{4 L_t L_s^3} \sum_{f=u,\,d,\,s} 
\Big\langle \tr \Big\{ \frac{\partial D^f_{\textnormal{st}}}{\partial b} 
{D^f_{\textnormal{st}}}^{-1}  \Big\} \Big\rangle\ ,
\label{Mdef}
\end{equation}
where $L_s$ and $L_t$ are the spatial and temporal extents of the lattice, measured in 
lattice spacing units. This observable has been measured over a 
few hundred thermalized trajectories for each parameter set and for 
each value of $b$, adopting a noisy estimator and averaging over 
40 different $Z_2$ random vectors for each single measure.

\begin{figure}[t!]
\includegraphics[width=0.9\columnwidth, clip]{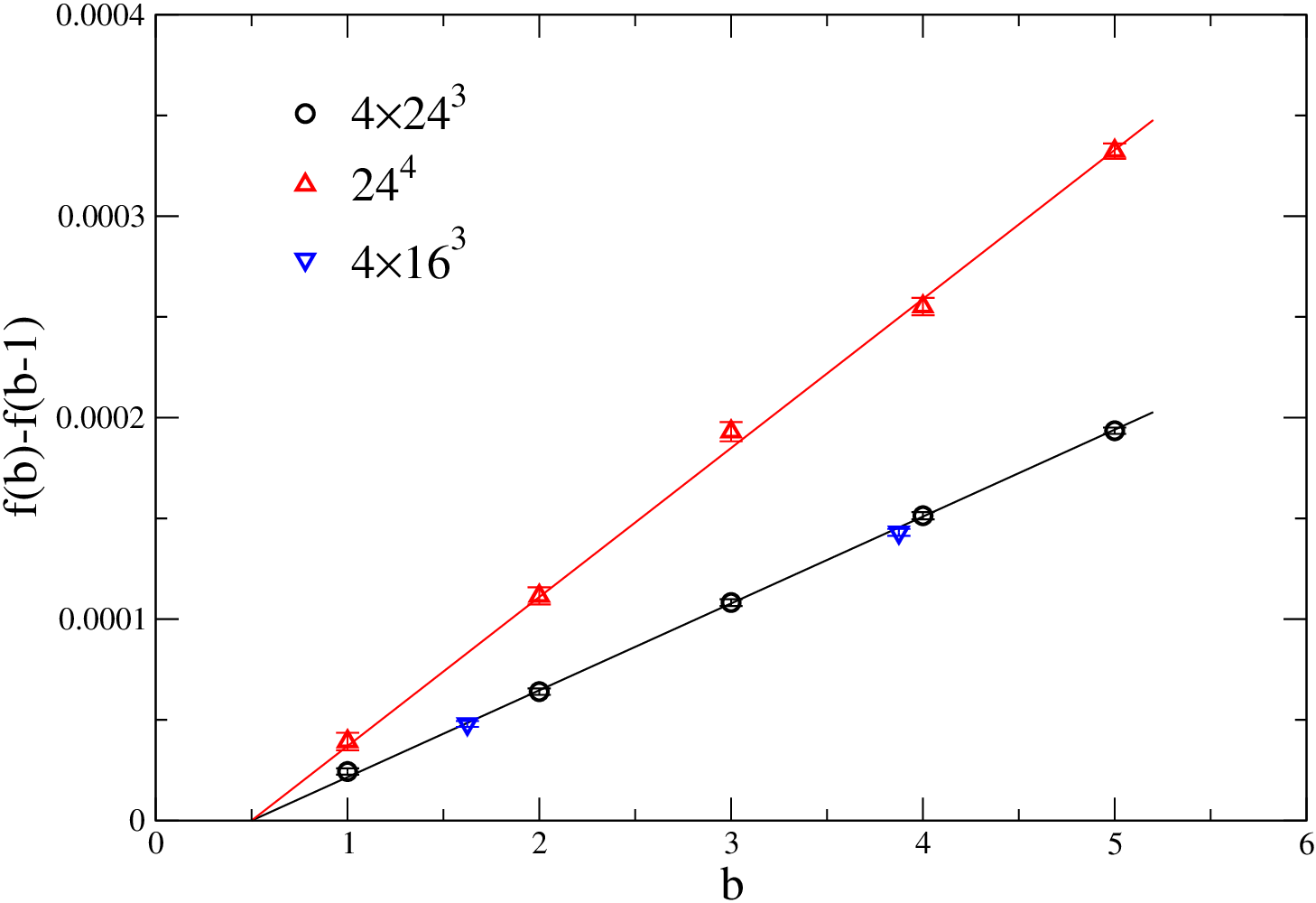}
\caption{$f(b) - f(b-1)$ computed for the same parameter
sets as in Fig.~\ref{M_fig}, together with best fits according
to Eq.~(\ref{dDF_fit}).
Two further, properly rescaled data points are reported
from a $16^3 \times 4$ lattice.}
\label{F_fig}
\end{figure}

\section{Numerical Results}
\label{results}

In Fig.~\ref{M_fig} we report an example of the determination
of the observable $M$, defined in Eq.~(\ref{Mdef}), for the largest
lattice spacing adopted and for two different lattice sizes,
$24^3 \times 4$ and $24^3 \times 24$: the former corresponds 
to $T \simeq 227$ MeV, the latter is our reference $T \sim 0$ lattice.
The range of explored values of $b$ spans the first
5 quanta of magnetic field and for each quantum we have determined
$M$ on a grid of $16$ equally spaced points. Such a grid
turns out to be fine enough to allow a reliable 
integration of $M$ (see Ref.~\cite{cffmm} for a 
discussion of the related systematic uncertainties).

The integral of $M$ over each quantum returns
the elementary finite differences $a^4 (f(b) - f(b-1))$.
Such quantities are more convenient than the whole
difference $\Delta f(b) = f(b) - f(0)$, since they can be 
determined independently of each other (one does not need to 
perform the whole integration from 0 to $b$) and have therefore 
independent statistical errors, thus allowing to exploit
standard fit procedures for uncorrelated data.

The finite differences corresponding to the data in Fig.~\ref{M_fig}
are reported in Fig.~\ref{F_fig}. We also report, after proper 
rescaling, data obtained
on a smaller $16^3 \times 4$ lattice and at the same value of the 
lattice spacing,
which show the absence of significant finite size effects.
Assuming that $a^4 \Delta f(b) \equiv c_2\, b^2 + O(b^4)$ holds
for integer $b$, then  
\begin{equation}
a^4\, (f(b) - f(b-1)) \simeq\, c_2\, (2b-1)\, .
\label{dDF_fit}
\end{equation}
Data in Fig.~\ref{F_fig} 
are well reproduced by such a behavior in the whole
explored range, and a number of different tests have been performed
to check the stability of our fit. In particular, the values obtained
for $c_2$ are stable, within errors, if the number of fitted points is 
changed, and also if a quartic term is added to the free energy,
i.e. if a fit of the form $\Delta f(b)=c_2b^2+c_4b^4$ is tried, which 
returns $c_2$ compatible, within errors, with the result obtained 
by the simple quadratic fit,
and $c_4$ compatible with zero. Finally, we have also tried 
a fit according to a generic power law behavior,
$\Delta f(b) \propto b^\alpha$, which returns $\alpha = 2$ within 
the precision of $1\%$.

It is interesting to notice that this implies that strongly interacting
matter behaves like a material with a linear response, at least
for magnetic fields up to $e B \sim 0.1$ GeV$^2$, corresponding to the 
highest field in the figure. Good
linear fits are obtained in similar ranges
of $eB$ for all explored values of $a$ and $T$.

\begin{figure}[t!]
\includegraphics[width=0.9\columnwidth, clip]{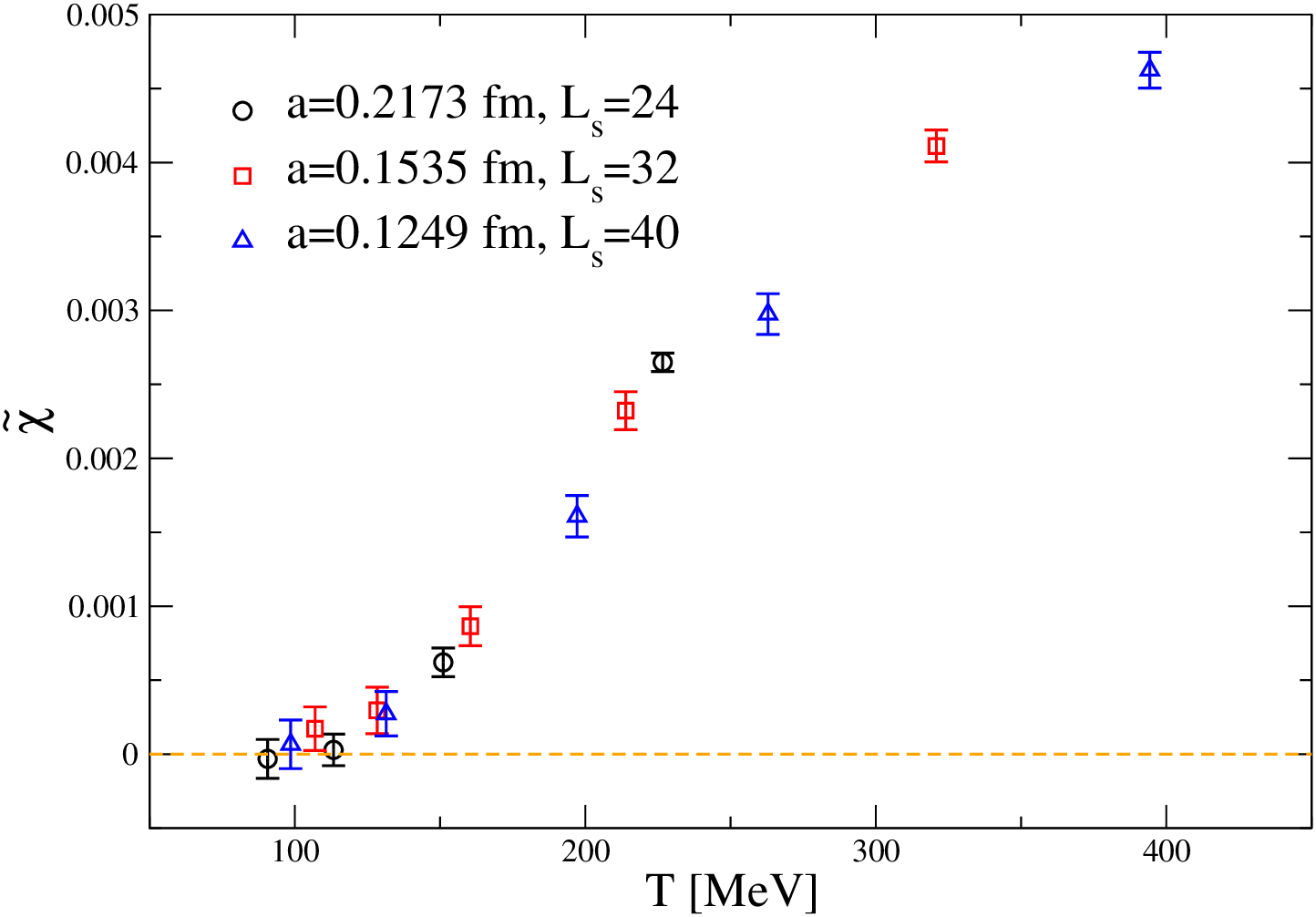}
\caption{
Susceptibility 
(SI units) as a function of $T$, for different values of the 
lattice spacing.}
\label{final_fig}
\end{figure}

The difference of the two slopes, 
${c_2}_R = c_2(L_t = 4) - c_2(L_t = 24)$, finally yields
the renormalized free energy
$a^4 \Delta f_R = {c_2}_R b^2 + O(b^4)$.
The determination of $\tilde \chi$ just 
requires a conversion into physical units for $\Delta f_R$ 
and $b$, according to Eq.~(\ref{bquant}). The result is 
\begin{equation}
\tilde \chi = - \frac{|e|^2 \mu_0 c}{18 \hbar \pi^2}\, L_s^4\, {c_2}_R \, ,
\end{equation}
in SI units ($\hbar$ and $c$ have been reintroduced explicitly). 
Instead, adopting natural units,
one obtains~\footnote{Actually, 
$\tilde\chi$ and $\hat\chi$, which are both dimensionless quantities,
are related to each other by a simple constant factor, 
$\hat\chi \simeq 10.9\ \tilde\chi$.}
(see Eq.~(\ref{intfree3}))
\begin{equation}
\hat\chi = - \frac{L_s^4\, {c_2}_R}{18 \pi^2}\ .
\end{equation}
A similar procedure has been repeated for all 
combinations of $T$ and $a$ reported in  
Table~\ref{param}. Results are shown in Table~\ref{tab2}
and in Fig.~\ref{final_fig}.

\begin{table}[t]
\centering
\begin{tabular}{ |c|c|c|c|c|c| }
\hline
$L_s$ & $L_t$ & a (\textrm{fm}) & $T (\textnormal{MeV})$ & \rule{0mm}{3.5mm} $10^3\tilde{\chi}$ & $10^2\hat{\chi}$  \\
\hline
24 & 4  & 0.2173(4) & 226(5)  &  2.648(62) & 2.887(68) \\ \hline
24 & 6  & 0.2173(4) & 151(3)  &  0.620(96) & 0.67(10)  \\ \hline
24 & 8  & 0.2173(4) & 113(2)  &  0.03(10)  & 0.03(11)  \\ \hline
24 & 10 & 0.2173(4) &  90(2)  &  -0.02(13) & -0.02(14)  \\ \hline
32 & 4  & 0.1535(3) & 321(6)  &  4.11(10)  & 4.48(11)  \\ \hline
32 & 6  & 0.1535(3) & 214(4)  &  2.32(12)  & 2.53(13)  \\ \hline
32 & 8  & 0.1535(3) & 160(3)  &  0.86(13)  & 0.94(14)  \\ \hline
32 & 10 & 0.1535(3) & 128(2)  &  0.29(15)  & 0.32(17)  \\ \hline
32 & 12 & 0.1535(3) & 107(2)  &  0.17(15)  & 0.19(16)  \\ \hline   
40 & 4  & 0.1249(3) & 394(8)  &  4.62(12)  & 5.04(13)  \\ \hline  
40 & 6  & 0.1249(3) & 263(5)  &  2.97(14)  & 3.24(15)  \\ \hline  
40 & 8  & 0.1249(3) & 197(4)  &  1.61(14)  & 1.75(15)  \\ \hline  
40 & 12 & 0.1249(3) & 131(3)  &  0.27(15)  & 0.30(16)  \\ \hline  
40 & 16 & 0.1249(3) & 99(2)   &  0.07(16)  & 0.07(18)  \\ \hline  
\end{tabular}
\caption{
Lattice parameters and results for $\tilde \chi$ and $\hat \chi$ (for the systematical error of the lattice
spacing see Tab.~(\ref{param})).}
\label{tab2}
\end{table}

\begin{figure}[t!]
\includegraphics[width=0.9\columnwidth, clip]{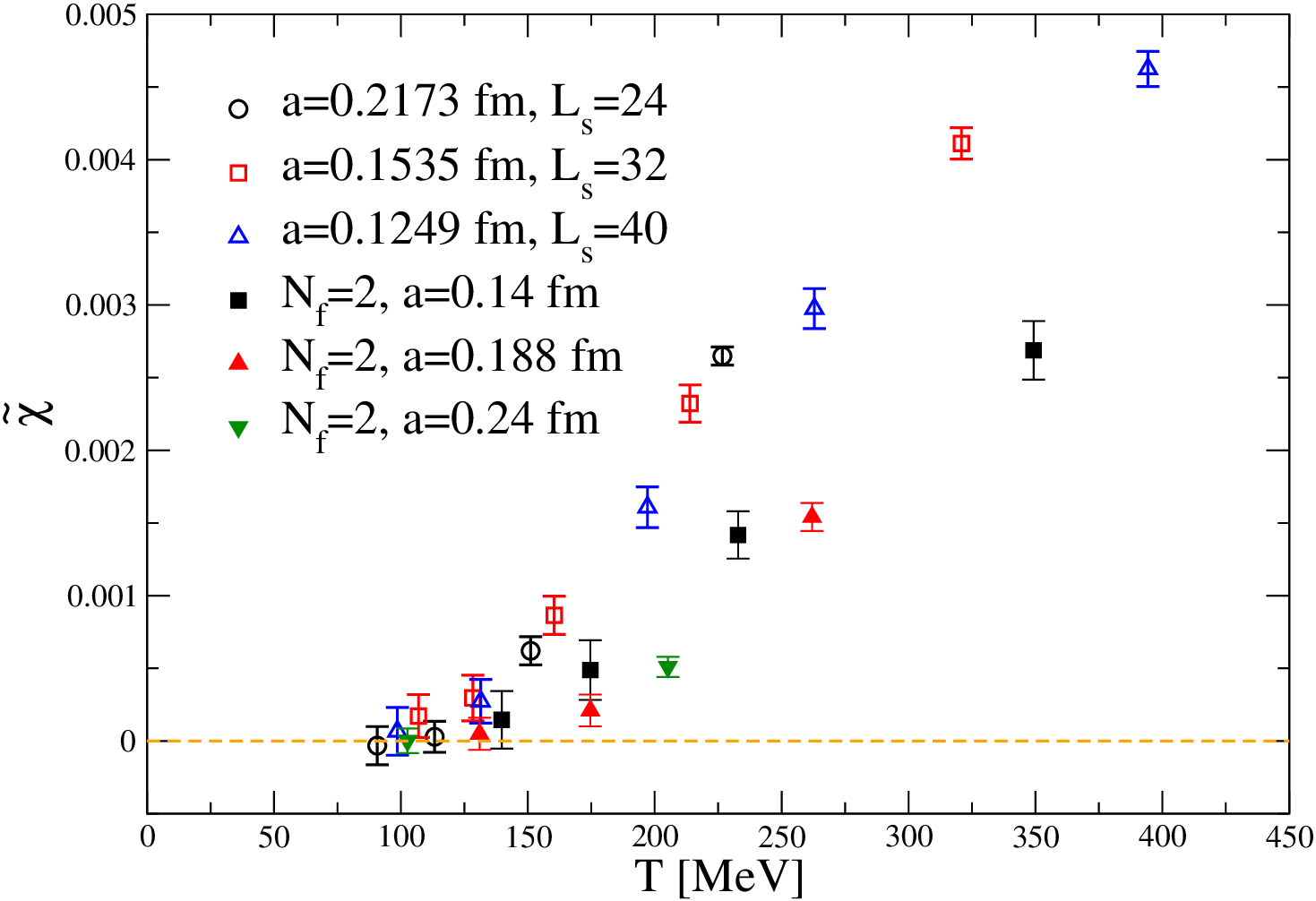}
\caption{Comparison of the results obtained for 
$N_f=2+1$ QCD at the physical point with those for 
$N_f=2$, $m_{\pi}=480\,\mathrm{MeV}$~\cite{cffmm}.}
\label{nf2_fig}
\end{figure}

\subsection{Discussion}
\label{discussion}

Data displayed in Fig.~\ref{final_fig} reveal various interesting 
features. First of all, one notices that the approach to the continuum
limit is very rapid and no significant
differences are observed between data computed at different
values of the UV cutoff.

\begin{figure}[t!]
\includegraphics[width=0.9\columnwidth, clip]{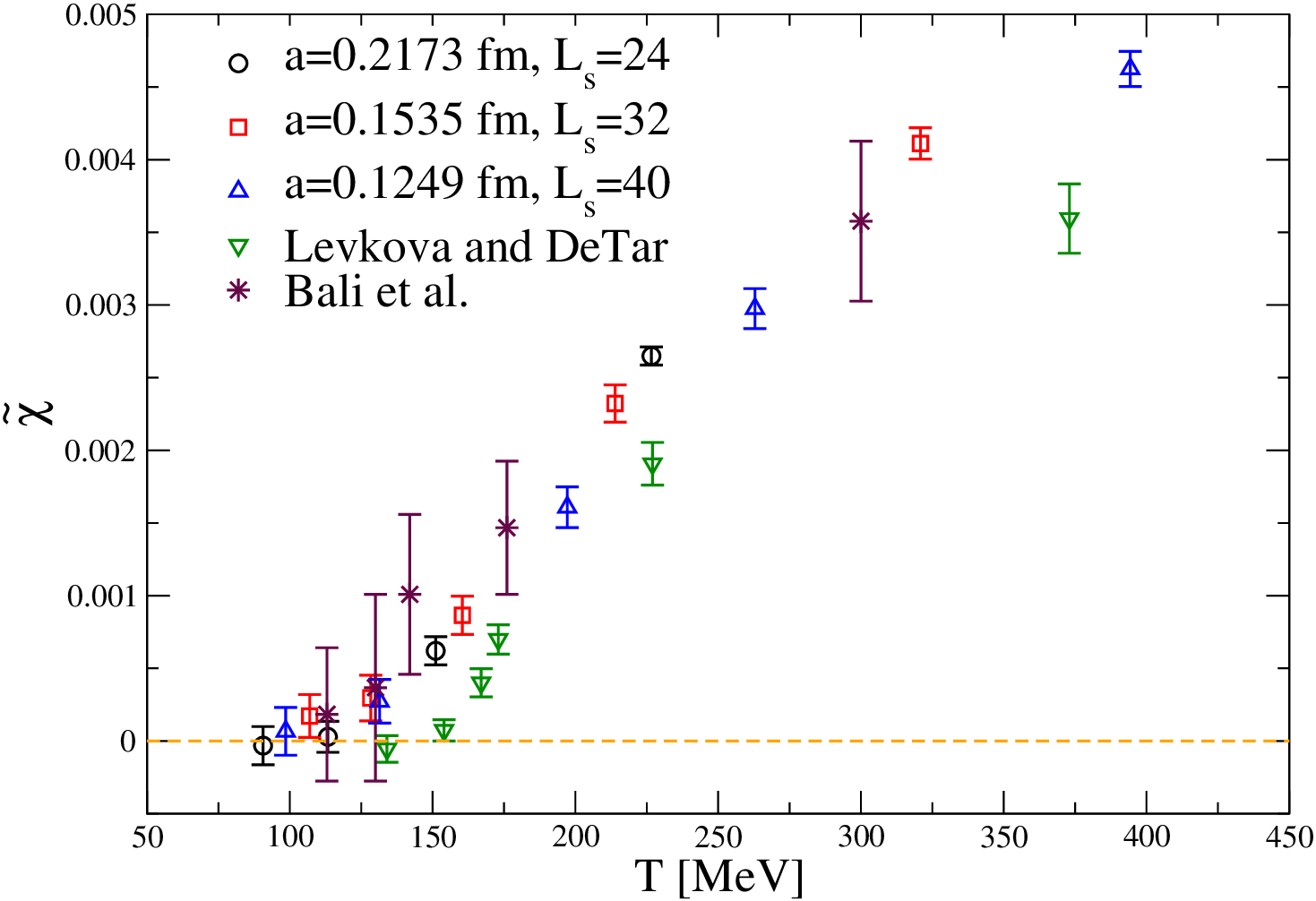}
\caption{Comparison of our results with those from 
Ref.~\cite{levkova} (Levkova and DeTar) and 
Ref.~\cite{reg4} (Bali {\it et al.}).}
\label{compare_others_fig}
\end{figure}

Results are in qualitative 
agreement with those of Ref.~\cite{cffmm}, obtained by exploiting the same 
computational method but for $N_f = 2$ with unphysical quark masses, see Fig.~\ref{nf2_fig}. Quantitative differences can be explained in part by the 
presence of the strange quark (which however gives a contribution 
to the total signal which is not larger than 1/6, see later), and
in part by the different pion mass: indeed, an increasing behavior
of $\tilde \chi$ with decreasing $m_\pi$ was already observed in 
Ref.~\cite{cffmm}.

We thus confirm that strongly interacting matter is paramagnetic, with a magnetic 
susceptibility which steeply rises crossing the deconfinement
transition, which is located around 
$150-160$ MeV~\cite{tchot,tcwup0,tcwup1,tcwup2}.
However, our data permit to better specify the behavior of the
magnetic susceptibility both in the low- and in the high-$T$ region,
and in particular around $T_c$.

In particular the magnetic susceptibility
is non-zero and positive, even if relatively smaller, 
also below $T_c$, and seems to vanish, within errors, 
for $T$ as low as 100 MeV. It is interesting to notice that data in the low
$T$ region can be fitted by a simple ansatz
$\tilde\chi = A\ \exp(-M/T)$, obtaining~\footnote{In order to check
for possible systematic effects related to the choice of the zero temperature
subtraction point, which in our case is set to $T = 40$ MeV, we 
have tried to repeat the fit by a function
$\tilde\chi = A\ (\exp(-M/T) - \exp(-M/T_0))$ with $T_0 = 40$~MeV,
but no differences, within numerical precision, have been appreciable,
as expected from the fact that the fit returns $M \gg T_0$.}, 
for $T < 170$ MeV, $M=870\pm 260$ MeV, 
hence in the correct ballpark of the lightest hadrons carrying 
a non-trivial magnetic moment (starting from the $\rho$ mesons), 
which are naturally expected to bring the major contribution
to the magnetic susceptibility in the hadronic phase. 
Therefore, this result seems in line with a Hadron Resonance Gas 
(HRG) model expectation; however, a direct comparison with 
the HRG prediction of Ref.~\cite{endrodihrg} is more easily
performed in terms of the magnetic contribution to the pressure
of the system, and will be presented at the end of this Section.

In Fig.~\ref{compare_others_fig} we report a comparison
with other existing lattice determinations 
of $\tilde \chi$ for $N_f = 2+1$ QCD.
The results of Ref.~\cite{reg4}, which have been obtained 
by the same lattice discretization of $N_f = 2+1$ QCD 
but exploiting a different method,
are in quantitative agreement with ours.
Agreement is found also with results
reported in Ref.~\cite{levkova}, apart from a limited region
below the transition, see Fig.~\ref{compare_others_fig}.
We do not think this discrepancy to be severe: 
it should be reconsidered after continuum extrapolation; moreover,
in the confined region, the finite size effects associated with 
the interface in the magnetic field could be more relevant, because
of the larger correlation lengths.
\\

\begin{figure}[t!]
\includegraphics[width=0.9\columnwidth, clip]{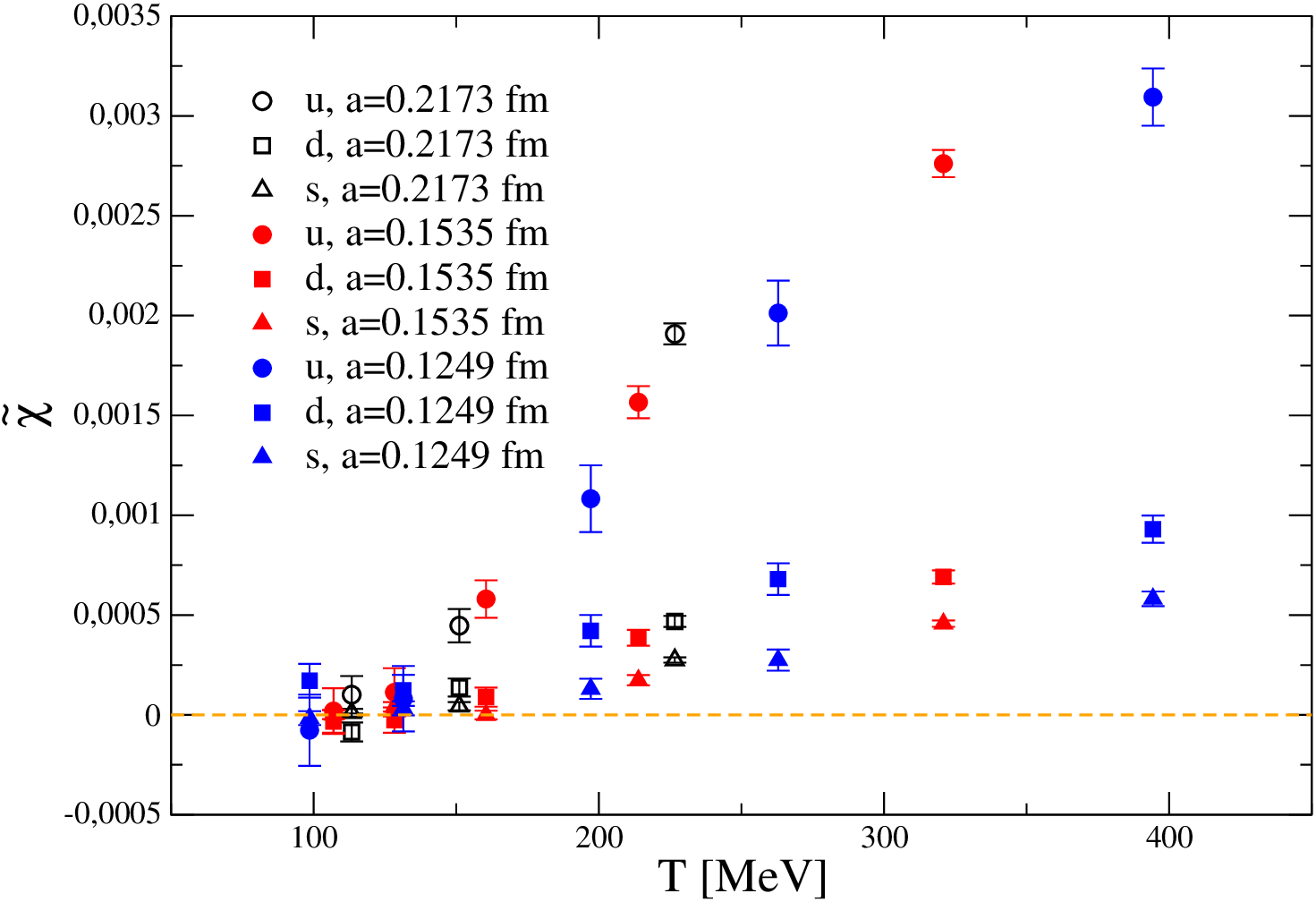}
\caption{Contributions to the susceptibility $\tilde\chi$ 
coming from the different flavors (see text for details).}
\label{flavours}
\end{figure}

\begin{figure}[t!]
\includegraphics[width=0.9\columnwidth, clip]{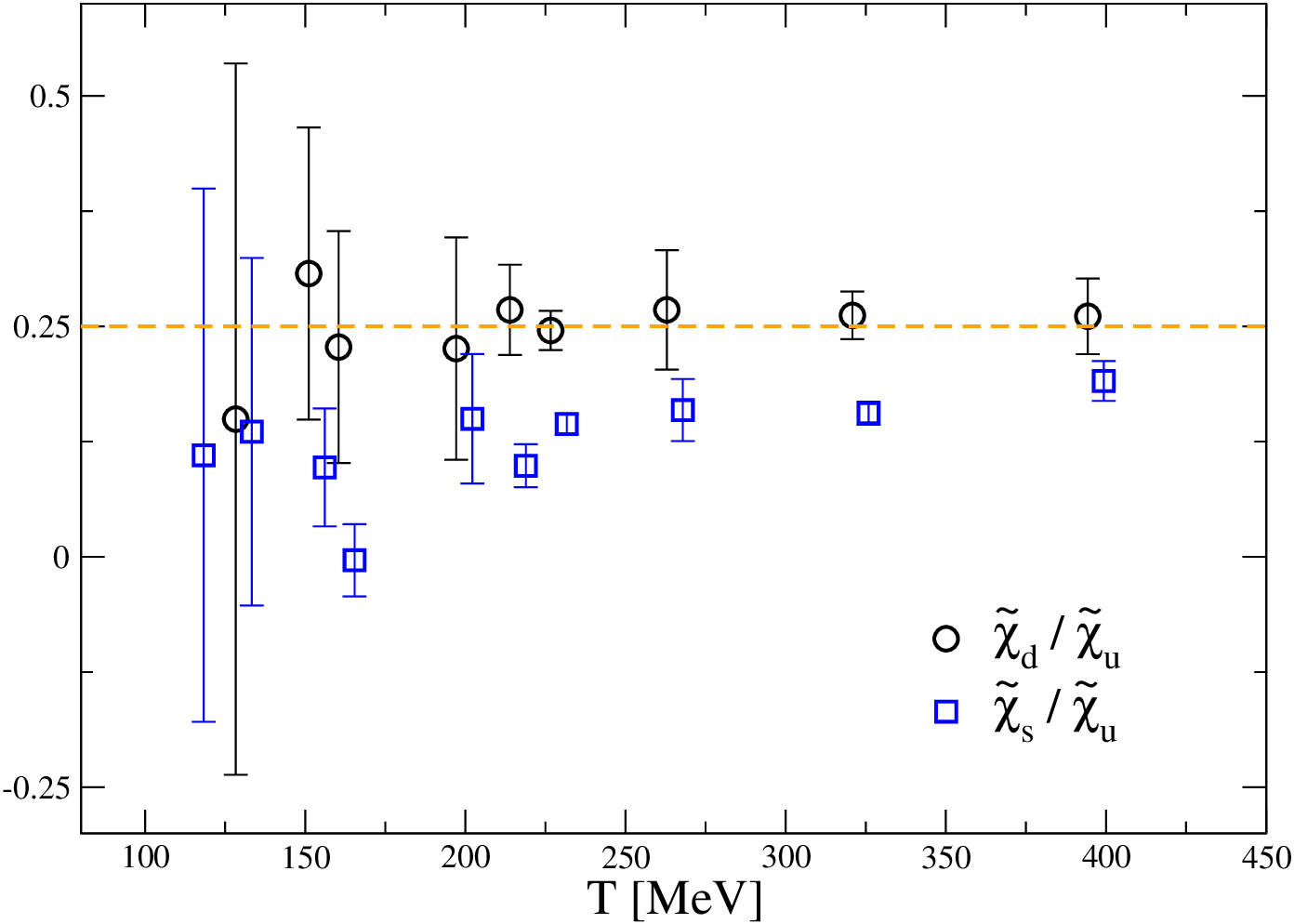}
\caption{Ratios of the flavor contributions
to the magnetic susceptibility as a function of $T$. 
No distinction is made in this figure for data corresponding
to different lattice spacings.
Data for $\tilde\chi_s/\tilde\chi_u$ have been 
shifted by 5 MeV along the $T$ axis to improve readability.} 
\label{flavratio}
\end{figure}

It is interesting to try disentangling the contributions to the magnetic
susceptibility coming from the different flavors. 
The observable that we integrate to reconstruct the free 
energy is naturally written as the sum of three different
contributions, $M = M_u + M_d + M_s$, see Eq.~(\ref{Mdef}),
and it is therefore straightforward to perform the analysis for the 
three different pieces in order to rewrite~\footnote{
We would like to stress that the separation
into different flavor contributions is not exact,
because of the mixings coming from quark loop contributions.}
$\tilde \chi = \tilde \chi_u + \tilde \chi_d + \tilde \chi_s$.
The corresponding quantities are reported separately in Fig.~\ref{flavours},
as a function of temperature. It is maybe more illuminating 
to look at the ratios $\tilde\chi_d/\tilde\chi_u$ and 
$\tilde\chi_s/\tilde\chi_u$, which 
are reported in Fig.~\ref{flavratio}. The former is in nice 
agreement, over the whole range of explored temperatures,
 with a leading order perturbative expectation based on the squared charge ratio, 
$(q_d/q_u)^2 = 0.25$. The latter, instead, seems to approach 
the same ratio from below in the high-$T$ limit: this is 
expected, since the thermal excitation of 
strange degrees of freedom is relatively 
suppressed because of the
higher quark mass.
\\

Let us now discuss the behavior of the magnetic susceptibility in 
the high temperature limit. Our data show an increasing behavior
over the whole explored range of temperatures, 
which however seems to flatten at the highest
values of $T$. It is interesting, in this respect, to compare 
our results with the lowest order perturbative prediction in the regime
of asymptotically high temperatures. Based on the results of 
Ref.~\cite{EPS}, in the high temperature limit the magnetic susceptibility is given by
\begin{equation}\label{pert_chi}
\hat{\chi}=\sum_f \frac{N_c q_f^2}{6\pi^2}\log(T/m_f)\ ,
\end{equation}
where $m_f$ and $q_f$ are the quark mass and the quark electric 
charge in units of $|e|$ and $N_c = 3$ is the number of colors.
From the physical point of view, even if the average magnetization 
of each single particle vanishes in the high temperature limit, because of thermal
disorder, the increase in the total density of 
thermally excited particles and antiparticles compensates that,
leading to a logarithmically diverging susceptibility.

\begin{figure}[t!]
\includegraphics[width=0.9\columnwidth, clip]{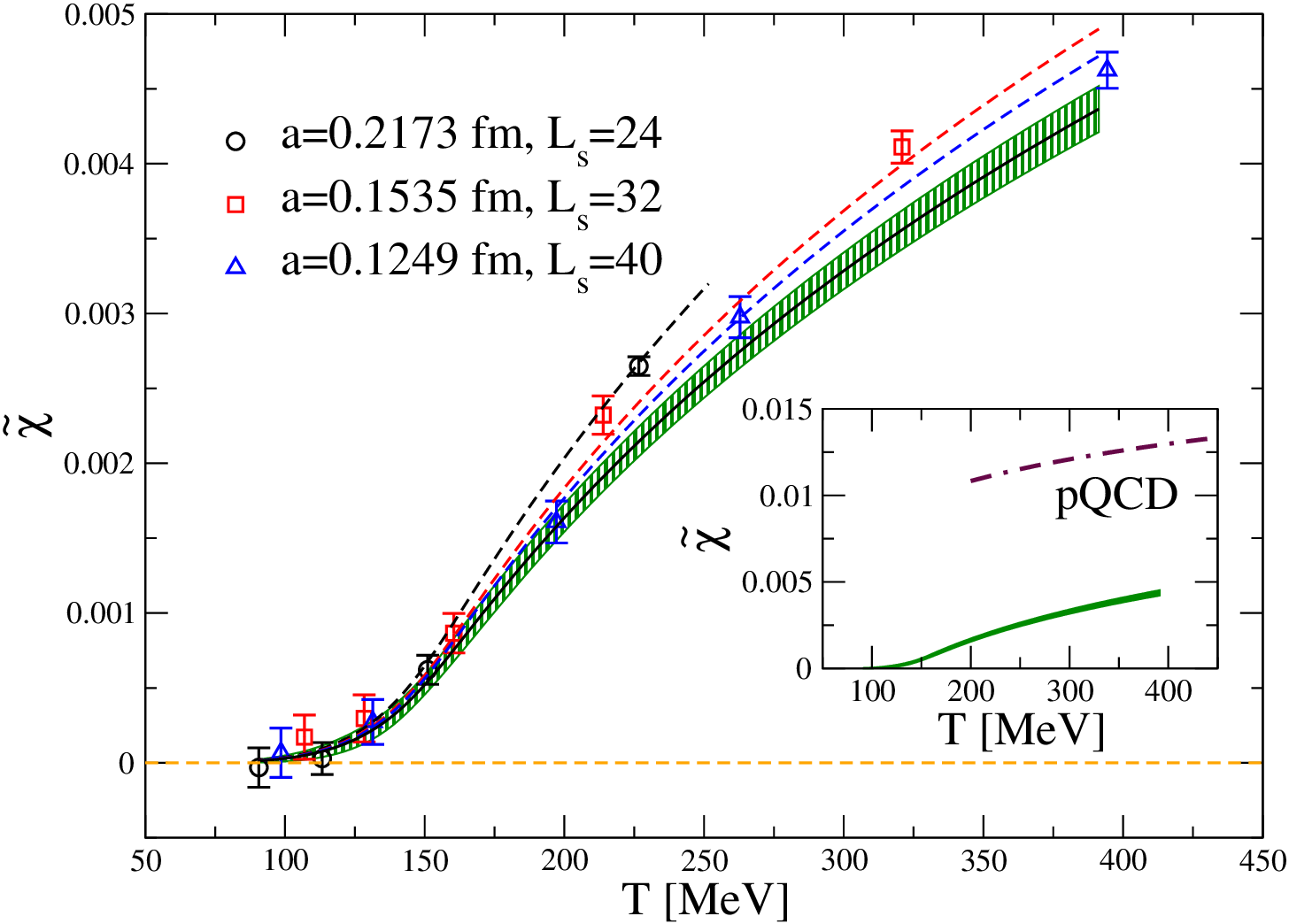}
\caption{
Continuum limit of the susceptibility (SI units) as a function of $T$: 
dashed lines are the fit result for the finite lattice spacing behaviours, the continuum extrapolation
and its error are indicated by the shaded region. 
The inset shows a comparison with the perturbative result
Eq.~\Eqref{pert_chi}.
}
\label{tot_fit}
\end{figure}

It is not reasonable to expect that
our data can be described by the free fermion result, however it is interesting to 
notice that $\tilde{\chi}(T)$ values in the 
temperature region $170\lesssim T \lesssim 400$ can be nicely fitted by a logarithmic 
behavior, although the coefficient is different from the perturbative one
(see Fig.~\ref{tot_fit} for a direct comparison of the perturbative 
estimate with our results).
\\

Regarding the continuum limit of our results,
the density of our data points does not permit a parametrization 
independent continuum extrapolation and requires instead to fix a definite
ansatz for the behavior of $\tilde \chi$.
Therefore, we tried to fit our data for the susceptibility in the whole 
temperature range by a function which reproduces the previously mentioned behaviors 
in the different regimes:
\begin{equation}
\tilde{\chi}(T)=\left\{
\begin{array}{ll}
A\ \exp(-M/T) & T\le \tilde{T} \\
A'\ \log(T/M') & T>\tilde{T} 
\label{eqconfit}
\end{array}\right.
\end{equation}
with a continuous and differentiable matching at the temperature $\tilde{T}$, 
which gives the constraints 
\begin{equation}
A' = A M \exp(- M/\tilde T)/\tilde T 
\quad
M' =  \tilde T \exp( -\tilde T/M)\ .
\end{equation}
Although we cannot neglect lattice artefacts (a fit to the whole data would give 
$\chi^2/\mathrm{d.o.f}\simeq 21.5/11$) our data are not dense and precise enough 
to extract the lattice spacing dependence of all the parameters. We explicitly 
verified that the two sets of fit parameters
\begin{itemize}
\item $\{\tilde{T}, A, M_0, M_2\}$ with $M=M_0+a^2M_2$
\item $\{\tilde{T}, A_0, A_2, M\}$ with $A=A_0+a^2A_2$
\end{itemize}
give equivalent results for the continuum extrapolation 
(with $\chi^2/\mathrm{d.o.f}\simeq 7/10$).
The value of $\tilde{T}$ turns out to be compatible with the transition 
temperature, $\tilde{T}=160(10)\,\mathrm{MeV}$, and the results of the fit 
are reported in Fig.~\ref{tot_fit} and in Appendix~\ref{contable} in table format.
One should stress that the errors reported for the continuum extrapolated 
values do not take into account the possible systematic error connected to 
the particular parametrization chosen for the extrapolation itself. 
That could
explain, in particular, the suppression of error bars in the low 
temperature region, where the parametrization is exponentially
suppressed, while we do not expect such effect to be significant
in the high temperature region.
\\

\begin{figure}[t!]
\includegraphics[width=0.9\columnwidth, clip]{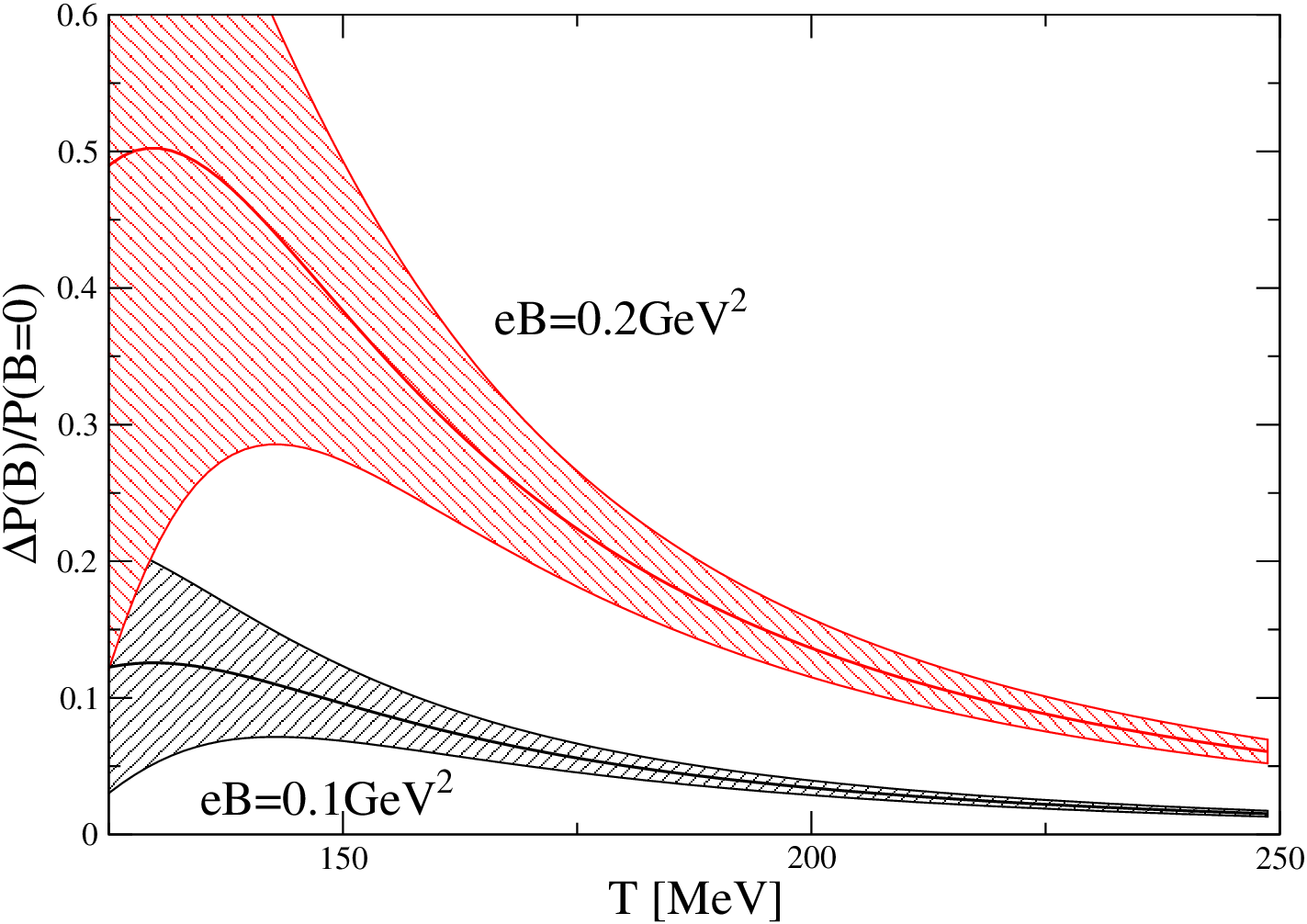}
\caption{Magnetic contribution to the pressure of strongly 
interacting matter, normalized to the pressure at zero 
magnetic field, as a function of temperature and for two
sample values of the magnetic field.} 
\label{pressure}
\end{figure}

Let us discuss now the effect of the magnetic field on the equation
of state of the system. The change in the pressure of the system
is easily obtained as
$\Delta P(B) = -\Delta f_R=\frac{1}{2}\hat{\chi} (eB)^2$ and is 
plotted in Fig.~\ref{pressure},
as a function of $T$, for two different values of the magnetic 
field (we make use of our continuum extrapolated determination), 
normalized by the pressure at $B = 0$ (data for the latter
quantity have been taken from Ref.~\cite{presref}). 
The different flavor contributions to $\Delta P(B)$ could be 
computed as well and, at the quadratic order in $eB$, they would 
be in the same relative ratios shown in Figs.~\ref{flavours} and
\ref{flavratio}.
We notice that the introduction of the magnetic field leads to a 
relative increase of the pressure 
which is larger around the phase transition,
and in the range 10-40\% for the typical fields produced in heavy ion
collisions at the LHC. In the high-$T$ regime, instead, the
relative increase rapidly approaches zero, as expected, since
the pressure at $B = 0$ diverges like $T^4$.

\begin{figure}[t!]
\includegraphics[width=0.95\columnwidth, clip]{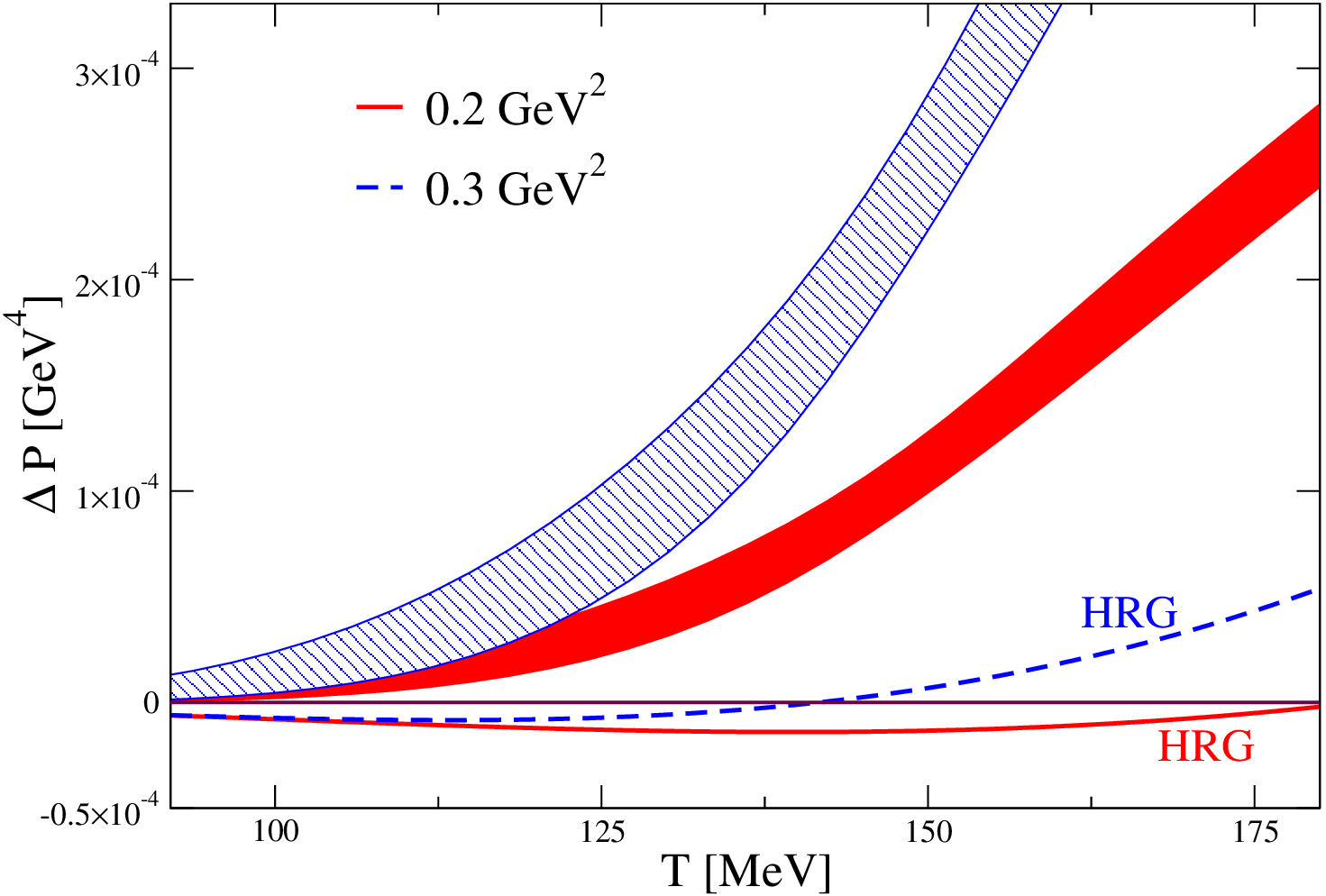}
\caption{Magnetic contribution to the pressure of strongly 
interacting matter, derived from our continuum extrapolated 
determination of the 
magnetic susceptibility, compared to the HRG prediction~\cite{endrodihrg}
for two different values of $eB$.}
\label{compare_hrg}
\end{figure}

Finally, in Fig.~\ref{compare_hrg}, we compare, for a couple
of values of $B$, 
$\Delta P(B) \simeq -\Delta f_R=\frac{1}{2}\tilde{\chi}B^2$,
computed from our continuum extrapolated 
susceptibility, with the (all orders) HRG prediction for the same 
quantity, after subtraction of vacuum contributions, extracted
from the results reported in Ref.~\cite{endrodihrg}. 
A few differences are clearly visible.
The HRG model predicts a slightly diamagnetic behavior for low
enough $T$, where the contribution from pions, which
is indeed diamagnetic, dominates: {notice that this behavior
of the HRG free energy,
which was not underlined by previous literature on the model,
regards only the thermal contribution,
i.e. after subtracting vacuum contributions to the free energy.
For larger temperatures or fields, 
instead, the contribution of higher spin hadrons becomes overwhelming
and the free energy behavior becomes paramagnetic.  
Lattice results do not confirm this possible weak diamagnetic behavior
for small temperatures and fields, even if, for $T \sim 100$ MeV,
the possibility is still open, within present statistical uncertainties.}

\section{Conclusions}
\label{summary}

We have investigated the magnetic properties
of strongly interacting matter at thermal equilibrium.
The study has been based on a lattice discretization of 
$N_f = 2+1$ QCD with physical quark masses: we have
considered 3 different values of the lattice spacing
and verified the absence of significant discretization effects.
We have exploited the method developed
in Ref.~\cite{cffmm}, which is 
based on a direct computation of the 
free energy density as a function of a 
uniform external magnetic background.

We confirm that the strongly interacting medium is paramagnetic,
both below and above the deconfinement transition, with a magnetic
susceptibility whose order of magnitude is comparable, within the 
explored range of temperatures, with those of well known 
strong paramagnetic materials, like liquid oxygen.
Moreover, data for the free energy density show that strongly
interacting matter behaves like a medium with a linear response, at least
for magnetic fields up to $e B \sim 0.1$ GeV$^2$, which is
the order of magnitude of the typical fields produced in 
non-central heavy ion collisions at the LHC; we stress 
that this  behavior is non trivial, since such fields cannot
be considered as small, when compared to the typical QCD scale
($0.1$ GeV$^2 \sim 5\, m_\pi^2$).

The magnetic susceptibility is relatively smaller in the confined
phase, and compatible with zero within present errors for temperatures
$T \lesssim 100$ MeV. It steeply rises across the deconfinement transition,
while in the high $T$ regime present data, which go up to $400$ MeV,
are compatible with a diverging logarithmic behavior, which is %also
predicted  by a 1-loop calculation of the free energy density.

Notice that the weak diamagnetism, which would be 
expected at low temperatures and fields based on a pion gas
approximation, and which is indeed predicted by HRG
computations~\cite{endrodihrg} (see also Ref.~\cite{Steinert:2013fza}),
is not confirmed by lattice data, even if present uncertainties
still leave room for it for $T \sim 100$ MeV. It would be interesting,
in the future, to further investigate this issue.

When compared to the pressure of the standard thermal medium, 
one sees that the introduction of an external magnetic field leads to 
a relative pressure increase which rapidly converges to zero
in the high-$T$ phase, while instead it 
gets larger around
the phase transition, where it 
is in the range 10-40\% for $e B$  0.1-0.2 GeV$^2$, i.e. for the typical 
fields produced in heavy-ion collisions. Such contribution becomes
rapidly larger and of $O(1)$ as the magnetic field increases, so that 
models for the cosmological QCD phase transition should 
necessarily take it into account.

We have computed the contributions to the magnetic susceptibility
coming from the different flavors: the $d$ contribution
is approximately $1/4$ of the $u$ contribution, as expected on 
a charge counting basis, over the whole range of explored temperatures;
the $s$ contribution is slightly smaller than the one of the $d$ flavor,
but tends to approach it in the high-$T$ limit.

Future studies, apart from extending the range of explored temperatures
and from increasing the precision of data in the low-$T$ regime,
should consider the 
effects of the inclusion of the $c$ quark, whose contribution,
especially in the high-$T$ region, could be comparable to those
from the $s$ and $d$ quarks, since the thermal suppression factor, due 
to the higher quark mass, should be partially compensated
by the electric charge factor. Considering the 
effects of the inclusion of a baryon chemical potential
might be interesting as well.

Finally, we would like to stress that, while present results provide
evidence that strongly interacting matter behaves like a medium
with a linear response for magnetic fields up to 
$e B \sim $ 0.1 GeV$^2$, non-linear corrections could be important
for larger fields, which may be of cosmological interest. 
Our method permits to reconstruct the full dependence of the 
free energy density on $B$ in a straightforward way: while the present
study was restricted on purpose to the linear response region,
in order to extract the magnetic susceptibility, we plan in the 
future to extend our analysis to larger fields, so as to 
determine the non-linear contributions too.

\acknowledgments

We thank G. Endrodi and E. Fraga for useful discussions.
FS thanks Michael Kruse for useful discussions regarding
code optimization on the Blue Gene/Q machine.
FN acknowledges financial support from the EU under 
project Hadron Physics 3 (Grant Agreement n. 283286).
Numerical simulations have been performed on the 
Blue Gene/Q Fermi machine at CINECA, based on the 
agreement between INFN and CINECA (under INFN project PI12) 
and on the CSN4 cluster of the 
Scientific Computing Center at INFN-PISA.

\appendix

\section{Continuum data for $\tilde{\chi}$}
\label{contable}

In Table~\ref{tabcont} we report the continuum extrapolated values of $\tilde{\chi}(T)$, obtained according to the parametrization reported in Eq.~(\ref{eqconfit}),
for some representative temperatures in the range $90\ \mathrm{MeV}\le T\le 395\
\mathrm{MeV}$.

\begin{table}[b!]
\centering
\begin{tabular}{ |r|l||r|l| }
\hline
$T (\mathrm{fm})$ & \rule{0mm}{3mm}$10^3\tilde{\chi}$ & $T (\mathrm{fm})$ & $10^3\tilde{\chi}$  \\ \hline
90.0  &  0.0117(92) &	245.0 &  2.45(10)   \\ \hline
95.0  &  0.018(13)  &	250.0 &  2.53(10)   \\ \hline
100.0 &  0.028(18)  &	255.0 &  2.61(10)   \\ \hline
105.0 &  0.042(23)  &	260.0 &  2.69(10)   \\ \hline
110.0 &  0.060(30)  &	265.0 &  2.77(11)   \\ \hline
115.0 &  0.084(37)  &	270.0 &  2.84(11)   \\ \hline
120.0 &  0.115(44)  &	275.0 &  2.92(11)   \\ \hline
125.0 &  0.154(50)  &	280.0 &  2.99(11)   \\ \hline
130.0 &  0.203(55)  &	285.0 &  3.06(11)   \\ \hline
135.0 &  0.263(59)  &	290.0 &  3.13(12)   \\ \hline
140.0 &  0.335(60)  &	295.0 &  3.20(12)   \\ \hline
145.0 &  0.420(60)  &   300.0 &  3.27(12)   \\ \hline
150.0 &  0.518(63)  &	305.0 &  3.34(12)   \\ \hline
155.0 &  0.627(68)  &	310.0 &  3.40(12)   \\ \hline
160.0 &  0.743(74)  &	315.0 &  3.47(12)   \\ \hline
165.0 &  0.861(80)  &	320.0 &  3.53(13)   \\ \hline
170.0 &  0.978(84)  &	325.0 &  3.60(13)   \\ \hline
175.0 &  1.094(87)  &	330.0 &  3.66(13)   \\ \hline
180.0 &  1.207(88)  &	335.0 &  3.72(13)   \\ \hline
185.0 &  1.318(90)  &	340.0 &  3.78(13)   \\ \hline
190.0 &  1.426(91)  &	345.0 &  3.84(14)   \\ \hline
195.0 &  1.531(92)  &	350.0 &  3.90(14)   \\ \hline
200.0 &  1.633(92)  &	355.0 &  3.95(14)   \\ \hline
205.0 &  1.733(93)  &	360.0 &  4.01(14)   \\ \hline
210.0 &  1.831(94)  &	365.0 &  4.07(14)   \\ \hline
215.0 &  1.926(96)  &	370.0 &  4.12(14)   \\ \hline
220.0 &  2.019(97)  &	375.0 &  4.18(15)   \\ \hline
225.0 &  2.110(98)  &	380.0 &  4.23(15)   \\ \hline
230.0 &  2.19(10)   &	385.0 &  4.28(15)   \\ \hline
235.0 &  2.28(10)   &	390.0 &  4.33(15)   \\ \hline
240.0 &  2.37(10)   &	395.0 &  4.39(15)   \\ \hline
\end{tabular}
\caption{Continuum extrapolated $\tilde{\chi}$ values.}
\label{tabcont}
\end{table}

\end{document}